\pdfoutput=1

\documentclass[journal]{IEEEtran}
\usepackage{color}
\usepackage{threeparttable}
\usepackage{amsmath}
\usepackage{amssymb}
\usepackage[pdftex]{graphicx}

\ifCLASSINFOpdf
\else
   \usepackage{graphicx}
\fi
\begin{document}
\title{Ultra-sensitive SQUID systems for pulsed fields -- Degaussing superconducting pick-up coils}

\author{Eva~Al-Dabbagh~and~Jan-Hendrik~Storm~and~Rainer K\"orber
\thanks{E. Al-Dabbagh was with Physikalisch-Technische Bundesanstalt, 10587 Berlin, Germany. She is now with TU Braunschweig, Institut f\"ur Konstruktionstechnik, 38106 Braunschweig, Germany.\newline J.-H. Storm and R. K\"orber are with Physikalisch-Technische Bundesanstalt, 10587 Berlin, Germany (e-mail: rainer.koerber@ptb.de)}}

\markboth{1EP3-08}%
{Shell \MakeLowercase{\textit{et al.}}: Ultra-sensitive SQUID system for pulsed field applications -- Degaussing supercondcucting pick-up coils}

\IEEEoverridecommandlockouts
\IEEEpubid{\makebox[\columnwidth]{10.1109/TASC.2018.2797544~\copyright2018
IEEE \hfill} \hspace{\columnsep}\makebox[\columnwidth]{ }} 
\maketitle

\begin{abstract}
SQUID systems for ultra-low-field magnetic resonance (ULF MR) feature superconducting pick-up coils which must tolerate exposure to pulsed fields of up to 100 mT. Using type-II superconductor niobium (Nb) field distortions due to trapped vortices in the wire result. In addition, their rearrangement after quick removal of the pulsed field leads to excess low frequency noise which limits the signal-to-noise ratio. In contrast, type I superconductors, such as lead (Pb), do not exhibit vortices but form an intermediate state with the coexistence of normal and superconducting domains.
\par We measured the magnetization loops of superconducting wire samples of Nb and Pb together with their noise behavior after pulsed fields. Pb also exhibits significant excess low frequency noise once the wire has been driven into the intermediate state. To avoid this problem, we removed the field not abruptly but in a linearly decaying sinusoidal manner thereby degaussing the wire. After application of 57~mT, we found that Nb can be degaussed within at least 50~ms, the shortest time used in this study. Pb can also be degaussed, albeit within 100~ms and a more complex dependency on the degaussing parameters. After successful degaussing, negligible excess low frequency noise is observed.

\end{abstract}

\IEEEpeerreviewmaketitle

\section{Introduction}
\IEEEPARstart{N}{ovel} techniques in biomagnetism based on ultra-low-field magnetic resonance (ULF MR) using superconducting quantum interference devices (SQUIDs) are currently developed. Such systems usually comprise low-$T_{c}$ current sensor SQUIDs, inductively coupled to a superconducting pick-up coil, and deploy a strong polarizing pulse of up to 100~mT prior to MR signal detection to boost the sample magnetization~\cite{ULFNMR2014}. However, to establish methods such as neuronal current imaging (NCI) or the combination of magnetoencephalography (MEG) and ULF MRI requires a significant improvement of the signal-to-noise ratio (SNR)~\cite{Koerber2016}.

\par In order to attain a higher SNR, ultra-low noise SQUID systems featuring a white noise level of about 150~aT~Hz$^{-1/2}$ have been developed~\cite{Storm2017}. Alternatively, a stronger polarizing field can be used, for which, the pick-up coil wire must tolerate exposure to pulsed fields.  Currently, the pick-up coil is most commonly made from the type-II superconductor niobium (Nb) as it has a relatively high lower critical field $\mu_{0}H_{c1}$ of about 140 mT at 4.2 K in high purity samples~\cite{Finnemore1966}. Nb wire is also widely available and easy to handle. If the applied field exceeds $H_{c1}$, trapped flux in form of vortices in the superconducting wire will result. This leads to field distortions and consequently to detrimental line broadening~\cite{Hwang2014}. In addition, their rearrangement after quick removal of the pulsed field leads to a random telegraph signal which manifests itself as excess low frequency noise which limits the SNR~\cite{Luomahaara2011,Storm2016}.

\par The use of type-I superconductors as a possible material for pick-up coils in SQUID-based ULF MR was also investigated, however with some conflicting results concerning trapped flux. Hwang $et~al.$~\cite{Hwang2014} did not observe line-broadening in NMR-signals from water using lead (Pb) pick-up coils after applying pulsed fields up to 160~mT, well above the critical field of the wire of 50~mT at 4.2~K \cite{Hwang2015}. This leads to the conclusion that Pb does not trap flux and therefore renders it a potential alternative. In contrast, Matlashov $et~al.$~\cite{Matlashov2015b} observed significant $1/f$-noise after pulsed fields using type-I superconductor Tantalum (Ta) pick-up coils which could be eliminated by thermocycling the Ta coil above it's $T_{c}$. This can be taken as evidence for rearrangement of trapped flux.

\par In this work we evaluated Pb as a possible alternative and investigated its performance regarding excess low frequency noise. In addition, we took a different approach to avoid excess low frequency noise. We removed the field not abruptly but in a decaying sinusoidal manner thereby degaussing the wire. Very recently, this method was successfully used for de-fluxing SQUIDs which have been exposed to pulsed fields~\cite{Matlashov2017}.

\section{Methods}

\subsection{Magnetization measurements}

The wires, denoted Nb1 (Supercon, SPC 414) and Pb1 (Goodfellow, PB005111/1) with the specifications given in Tab.~\ref{tab:wire_samples}, were characterized by performing magnetization curve measurements at 4.2 K using a magnetic properties measurement system (MPMS, Quantum Design). The wire samples were 5~mm long and cooled in zero field.

\begin{table}[!t]
\renewcommand{\arraystretch}{1.3}
\caption{Specifications of used wire samples.}
\label{tab:wire_samples}
\centering
\begin{threeparttable}
\begin{tabular}{c c c c c c}
\hline \hline
	Wire	&	Supplier	&	Diameter	&  critical field 	&	Purity 	& Insulation\\
				&						& ($\mu$m)	& 			(mT) 											&  (\%) & 	\\
\hline
	Nb1 		& Supercon	& 	101.6 	& 			105$^{a}$ 											& 	99.96+$^{c}$ & polyimide-enamel		\\
	Pb1 		&	Goodfellow& 	250  		& 			50$^{b}$ 												& 99.95 & -	\\
\hline
\end{tabular}
\begin{tablenotes}[para,flushleft]
$^{a}$ $\mu_{0}H_{c1'}$\\
$^{b}$ $\mu_{0}H_{c'}$ \\
$^{c}$ representative value (commercial grade Nb)
\end{tablenotes}
\end{threeparttable}
\end{table}

\subsection{Test gradiometers of Nb and Pb}
Compact first order axial gradiometers were wound on a polyoxymethylene (POM) holder having a diameter of 41.5~mm and a baseline of 3.5~mm. The bare Pb wire was insulated using GE-Varnish. Each gradiometer was in turn connected to the same current sensor SQUID which was housed inside a Nb shielding. The single stage SQUID was equipped with on-chip current limiters and had an input coil inductance $L_{i}$ of 150~nH~\cite{Drung2007}. The probe was operated in our ultra-low noise dewar LINOD2~\cite{Storm2017} and the measurements carried out inside a 2-layer magnetically shielded room.

\subsection{Field applications -- Degaussing procedure}
\begin{figure}[!t]
\centering
\includegraphics[width=.70\columnwidth]{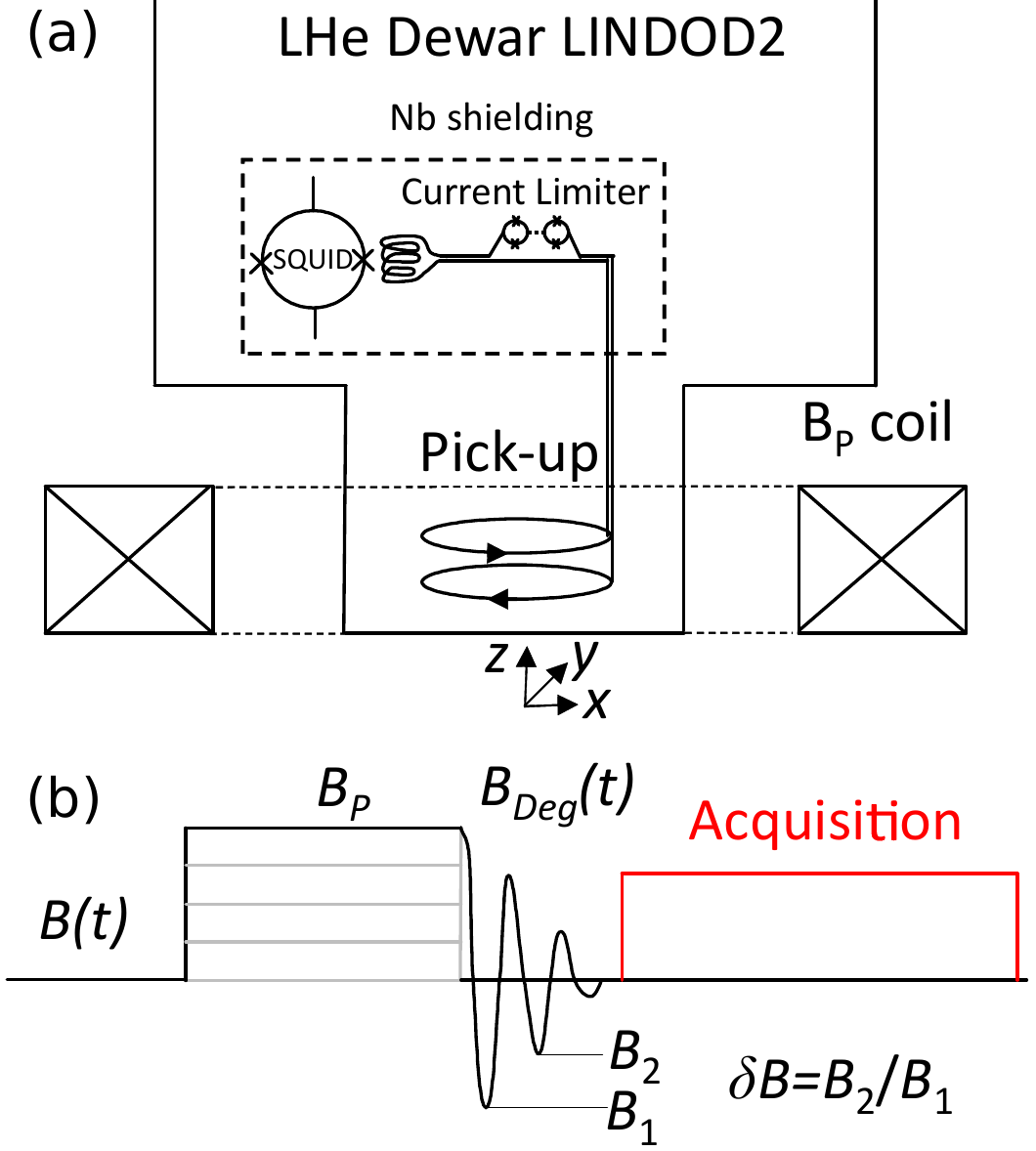}
\caption{(a) Schematic diagram of the experimental setup showing the arrangement of the polarizing coil and the gradiometers. (b) shows the sequence used for fast turn-off and degaussing with $\delta B$ being the fractional change for $B_{Deg}$ after one period.}

\label{fig:schematic_setup}
\end{figure}

The schematic experimental setup is shown in Fig.~\ref{fig:schematic_setup}. Successively larger magnetic fields $B_{P}$, starting from zero, were applied via a compact coaxial solenoidal room temperature coil with a central field current ratio of 2.15~mT/A. It was driven with currents of up to 25~A with a commercial power amplifier. The centers of the gradiometer and the polarizing coil coincided leading to a somewhat larger $B_{P}$ at the wire position with the field being perpendicular to the gradiometer loops. This leads to a demagnetization factor of close to two. A fast, linear turn-off of $B_{P}$ was achieved by discharging the coil via home-built electronics with a ramp of 20~kA/s. For a maximum applied field of 57~mT the turn-off time was about 1.25~ms. 

\par The degaussing function $B_{Deg}$ starting at the end of the polarizing field $B_{P}$ was a linearly decaying cosine function:
\begin{equation}
	B_{Deg}(t)=B_{P}(1-t/t_{Deg})\cos(2\pi f_{Deg}t)
	\label{eq:Degaussing}
\end{equation}
The length $t_{Deg}$ and the frequency $f_{Deg}$ were varied between 50 and 250 ms and 10 and 100 Hz, respectively. Larger values for $f_{Deg}$ could not be implemented due to the inductance of the polarizing coil limiting the output of the power amplifier during the initial phase of the degaussing procedure. Data were captured for 935~ms and analyzed 50~ms after turn-off of the magnetic field.

\par In order to assess the influence on the spin dynamics during the degaussing procedure we solved the Bloch equations numerically for $B_{Deg}$ with 50~ms and 70~Hz parallel and perpendicular to a detection field $B_{Det}$ of 38.64~$\mu$T. The latter case is of particular importance with regard to adiabatic or non-adiabatic turn-off.

\section{Results and discussion}

\subsection{Magnetization curves}

\begin{figure}[!t]
\centering
\includegraphics[width=.75\columnwidth]{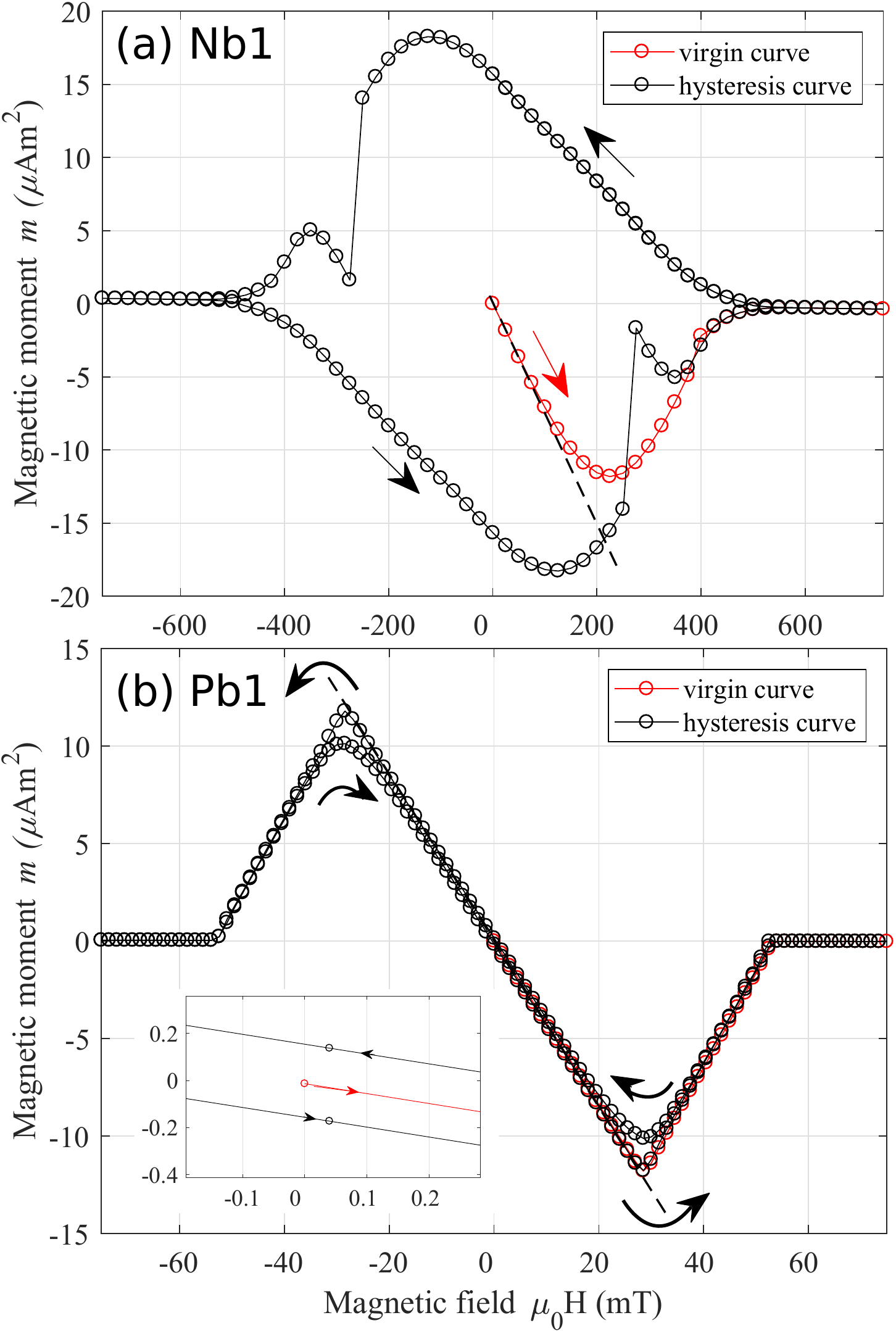}
\caption{Magnetization loops for a) Nb1 and b) Pb1 with the magnetic field perpendicular to the wire. The dashed line indicate perfect diamagnetism.}
\label{fig:hysteresis}
\end{figure}

In Fig.~\ref{fig:hysteresis} the magnetization curves of Nb1 and Pb1 for the field applied perpendicular to the wires are shown. From the virgin curves and by analyzing $\mu_{0}H/m$ vs. $\mu_{0}H$~\cite{Storm2016}, we extract the fields at which flux starts to penetrate into the wires as $\mu_{0}H_{c1'}=50$~mT for Nb1 and  $\mu_{0}H_{c'}=27$~mT for Pb1, respectively. The values given in Tab.~\ref{tab:wire_samples} are determined for parallel fields and hence negligible demagnetizing effects. 
\par A large hysteresis loop is observed for Nb1 indicating substantial flux trapping. The flux jumps in the magnetization curves at $\pm$250~mT are probably due to magnetothermal instabilities where abrupt flux entry causes a temperature increase driving most of the sample normal~\cite{Swartz1968}. In contrast, Pb1 shows only a minute hysteresis loop which can also be seen in the inset and exhibits the expected behavior for a cylinder in a perpendicular field. Above 27~mT the wire is in the intermediate state, in which normal and superconducting regions coexist.

\subsection{Noise after rapid turn-off}

\begin{figure}[!t]
\centering
\includegraphics[width=.80\columnwidth]{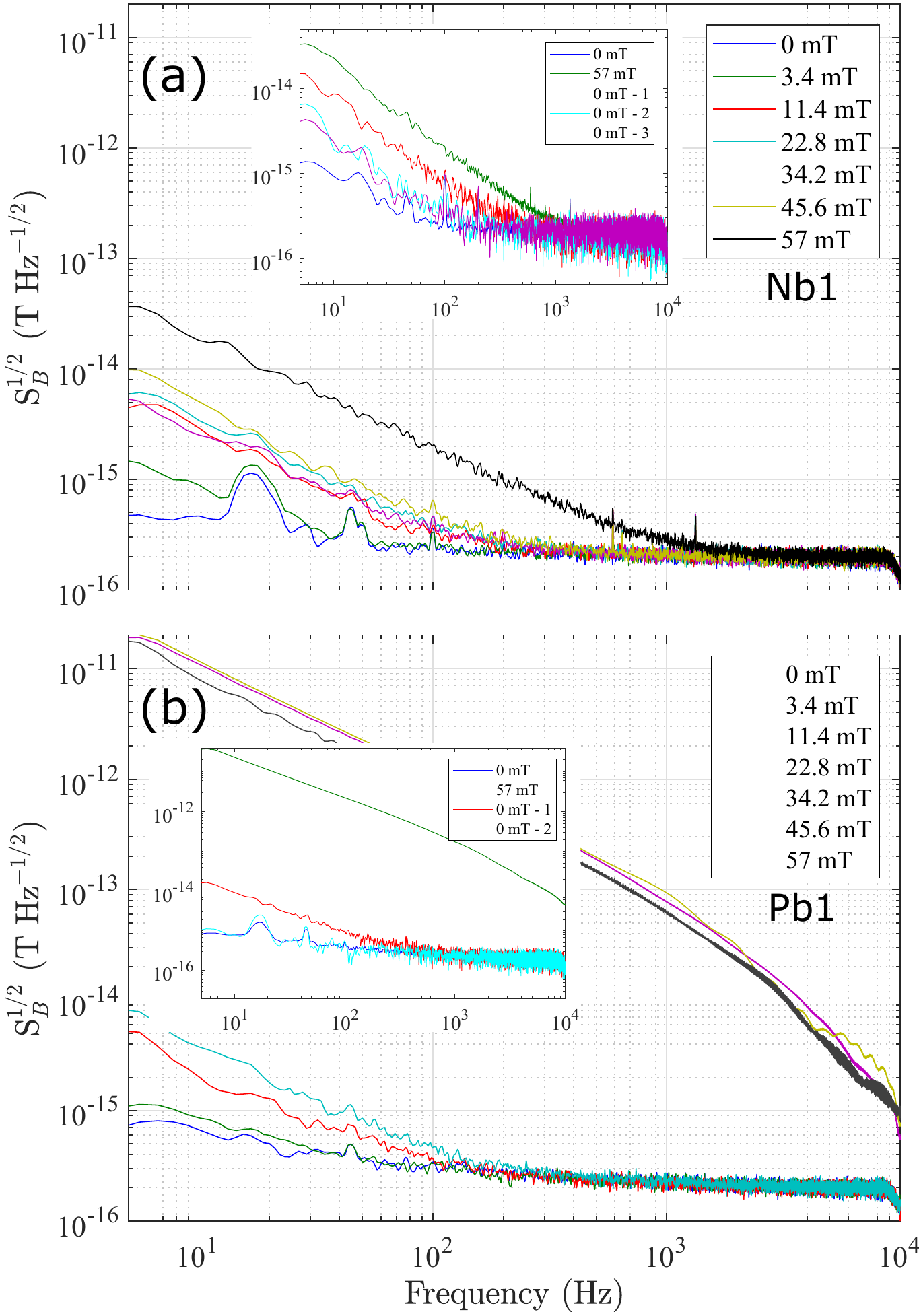}
\caption{$S_{B}^{1/2}$ after rapid turn off for (a) Nb1 and (b) Pb1. The insets show the relaxation behavior after a single pulse with 57~mT.}
\label{fig:LSD_rapid_turnoff}
\end{figure}

The flux density noise $S_{B}^{1/2}$ ($S_{B}$ being the power spectral density of the noise) after pulsing using the rapid turn-off is shown in Fig.~\ref{fig:LSD_rapid_turnoff} for gradiometers made from Nb1 and Pb1. We quote the field experienced by the wire. 

A threshold behavior is observed in Nb1 as reported before~\cite{Storm2016}. Briefly, once the critical field $\mu_{0}H_{c1'}$ of 50~mT is exceeded, a significantly larger low frequency noise is observed. Below 50~mT, there is some increased noise of unknown origin. 

\par In comparison, Pb1 shows a similar excess low frequency noise for fields below 27~mT. Here, the wire is in the Meissner state for this geometry as can be seen in Fig.~\ref{fig:hysteresis} (b). The excess noise increases somewhat with increasing pulsed field amplitudes in this range. Then, for fields above 27~mT when Pb1 is in the intermediate state, the behavior is identical for all fields and shows markedly increased noise. This is mainly due to flux jumps in the SQUID signal which are induced by rearrangement of flux within the Pb wire causing a signal change above the slew rate. In the intermediate state in type-I superconductors normal and superconducting regions coexist throughout the sample, pinning the internal field to $H_{c}$. Therefore, the enormous excess low frequency noise is already observed for fields just above 27~mT.

\par The insets in Fig.~\ref{fig:LSD_rapid_turnoff} show the spectra after a single pulse with 57~mT followed by noise measurements taken every 12.5~s. For Nb1, the excess low frequency noise decays within about 25~s but remains somewhat higher than the reference noise. Hence, rearranging flux remains in the sample. For Pb1, the flux rearrangement also occurs within roughly 25~s but no excess low frequency noise and consequently flux rearrangement is observed thereafter. This shows that Pb tends to expel flux as this would constitutes an unstable thermodynamical state in the type-I superconductor.

\subsection{Noise after degaussing}

\begin{figure}[!t]
\centering
\includegraphics[width=0.80\columnwidth]{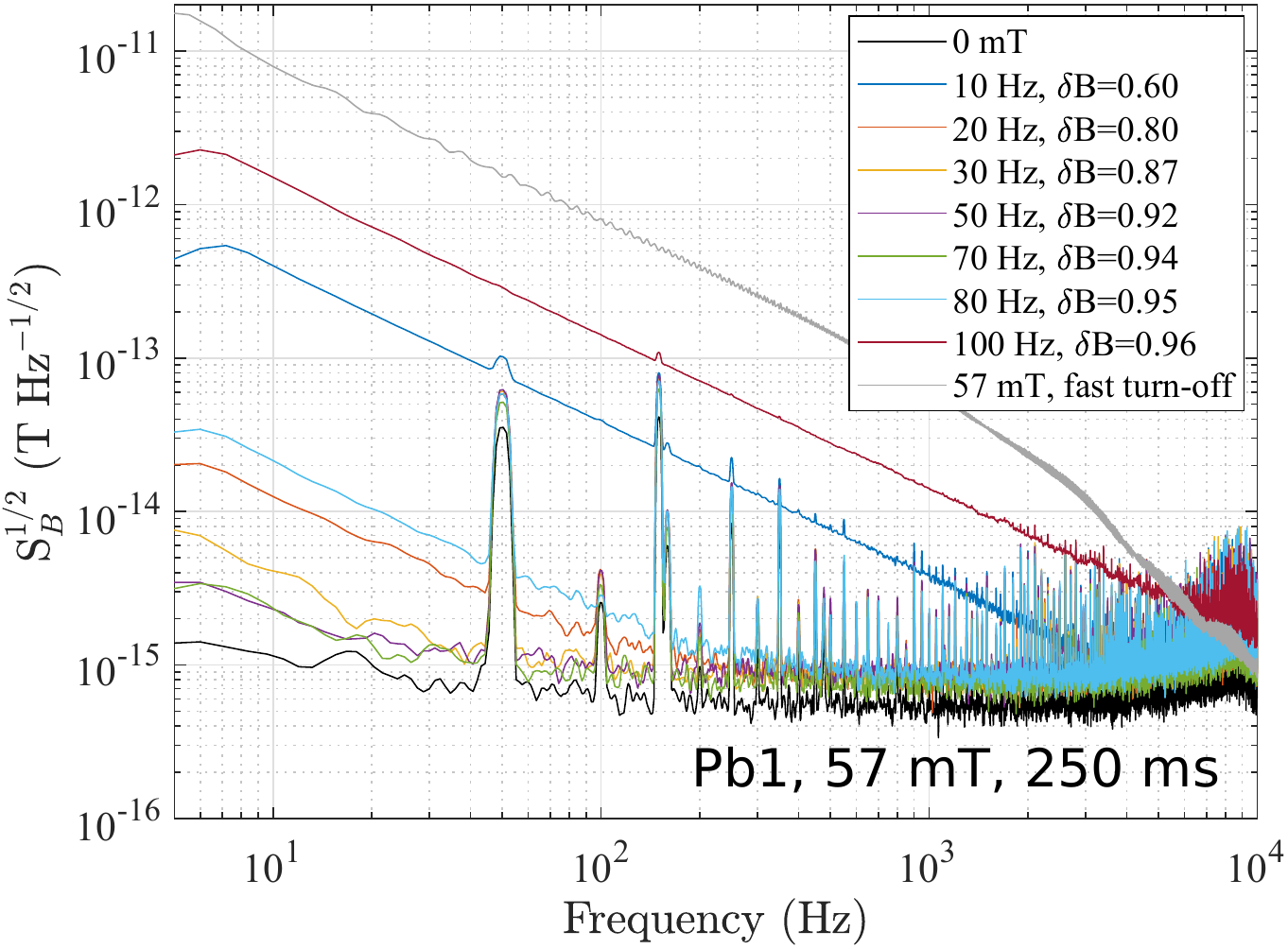}
\caption{$S_{B}^{1/2}$ after degaussing for different frequencies and a $t_{Deg}$ of 250~ms for Pb1. The power line interference and the increased white noise are due to a changed setup necessary to enable the application of a bipolar current. The gray trace shows the noise after fast turn-off for comparison.}
\label{fig:deg_frequency}
\end{figure}

The flux density noise $S_{B}^{1/2}$ after degaussing the Pb1 gradiometer using a constant degaussing time $t_{Deg}$ of 250~ms and a varying degaussing frequency $f_{Deg}$ from 10 to 100~Hz is shown in Fig~\ref{fig:deg_frequency}. The data obtained for Nb1 are not shown as there was no excess low frequency noise observed for the parameters used which leads us to conclude that Nb1 can be degaussed for all frequencies used within 250~ms. For Pb1 the results are more complex. Within the frequency range of 30 to 70~Hz, effective degaussing can be achieved leading to minimal excess low frequency noise. For smaller and larger frequencies degaussing is not as efficient. 

\par
\begin{figure}[!t]
\centering
\includegraphics[width=.80\columnwidth]{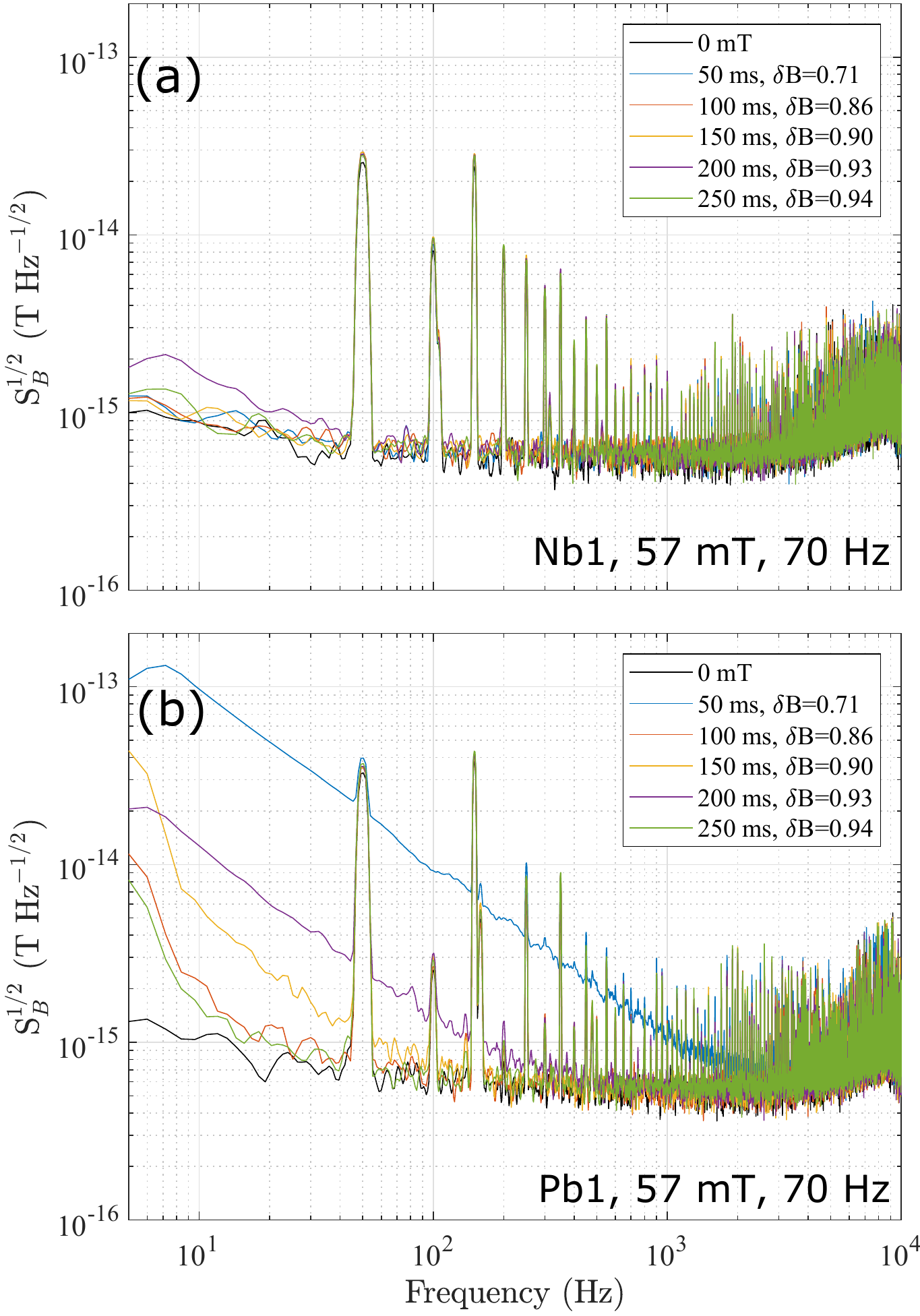}
\caption{$S_{B}^{1/2}$ after degaussing for different times and a $f_{Deg}$ of 70~Hz for (a) Nb and (b) Pb.}
\label{fig:deg_duration}
\end{figure}

 Based on the above results, we chose the constant degaussing frequency $f_{Deg}$ of 70~Hz and varied the degaussing time $t_{Deg}$ from 50 to 250~ms. $S_{B}^{1/2}$ is shown in Fig~\ref{fig:deg_duration}. Similar to before, Nb1 can be degaussed for $t_{Deg}$ as low as 50~ms. The absence of any increased excess low frequency noise for all degaussing parameters leads us to the conclusion that even shorter $t_{Deg}$ are possible. However, this assumption should be confirmed by further experiments. For Pb1 the situation is again more complex. Excess low frequency noise is minimal above 10~Hz for $t_{Deg}$ of 250, 150 and 100~ms. It appears, that times shorter than 100~ms are not suitable. Note, we do not rule out that for $t_{Deg}\neq 250$~ms a different $f_{Deg}$ might be optimal which should be addressed in a more detailed study.

\par As we have seen, the type-II superconductor Nb1 can be degaussed by a decaying AC-field for all parameters used. Matlashov \textit{et al.} proposed vortex-antivortex annihilation as as possible mechanism to explain the observed inductive de-fluxing effects in their experiments on thin film LTS-SQUIDs~\cite{Matlashov2017}. This process is also consistent with our experiments using bulk wire which was exposed to fields just above $H_{c1'}$ leading to vortices only at the surface. 

\par The origin for complex dependence on the degaussing parameters in Pb1 is presently unknown, but can to some degree be linked to $\delta B$, the fractional change of $B_{Deg}$ after one period. For $t_{Deg}=50$~ms, $f_{Deg}=70$~Hz and $t_{Deg}=250$~ms, $f_{Deg}=10$~Hz and 20~Hz, $\delta B$ is smaller than $\sim 0.8$ and effective degaussing cannot be achieved. Note, excess low frequency noise is also occasionally observed for parameters with $\delta B\gtrsim 0.8$, as for instance for $t_{Deg}=250$~ms and $f_{Deg}>70$~Hz (see Fig.~\ref{fig:deg_frequency}) or $t_{Deg}=200$~ms and $f_{Deg}=70$~Hz (see Fig.~\ref{fig:deg_duration} (b)). Hence, $\delta B\gtrsim 0.8$ seems to be merely a necessary condition rather than a sufficient one, and other, yet unknown, parameters are also important for the successful degaussing of Pb1.

\par The question as to whether the degaussing procedure removes all trapped flux remains unanswered. Strongly pinned flux would not rearrange and consequently not cause any excess low frequency noise. Line broadening in NMR experiments on samples with long relaxation times would reveal any field distortions due to permanently pinned flux.

\subsection{Influence on spin dynamics}

\begin{figure}[!t]
\centering
\includegraphics[width=.90\columnwidth]{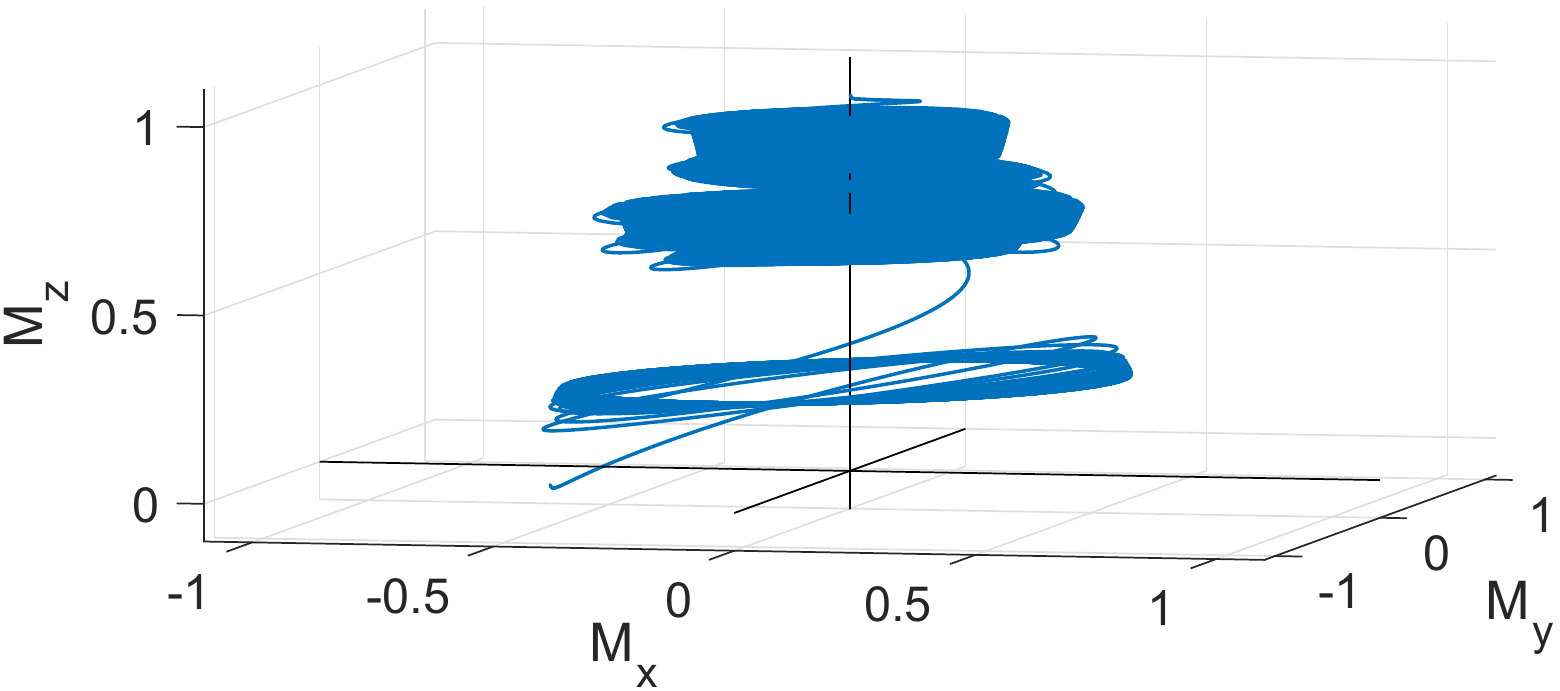}
\caption{Simulation of the magnetization trajectory for $B_{Deg}\perp B_{Det}$ during the degaussing procedure with $B_{Deg}(0)=53.8$~mT, $f_{Deg}=70$~Hz, $t_{Deg}=50$~ms and $B_{Det}=38.64~\mu$T. Relaxation was incorporated in the simulation assuming $T_{1}=T_{2}=100$~ms.}
\label{fig:spin_dynamics}
\end{figure}

If such a degaussing process is to be used for ULF MRI a sound knowledge on the spin evolution is required.
In Fig.~\ref{fig:spin_dynamics} the spin dynamics during the degaussing sequence with $B_{Deg}(0)=53.8$~mT is shown. This was calculated for a detection field of 38.64~$\mu$T, corresponding to a precession frequency of 1645~kHz, perpendicular to $B_{Deg}(t)$. There is a strong influence on the spin dynamics at each zero crossing of the degaussing field as the effective field $B_{eff}$ changes direction, and $dB_{eff}/dt$ determines whether the magnetization $M$ can follow $B_{eff}$. Since $dB_{eff}/dt$ at each zero crossing gets successively smaller these effects become more important. In our example, $M_{z}$ actually becomes negative towards the end. The final angle $\alpha$ spanned by $M$ and $B_{Det}$ is reduced from 90$^{\circ}$ to 9.15$^{\circ}$. Further simulations show, that $\alpha$ is very sensitive to $B_{Deg}(0)$, e.g. for 53~mT we found $\alpha=69.7^{\circ}$. Hence, a controlled non-adiabatic turn-off during the degaussing process might be difficult to achieve, as for instance due to inhomogeneities in $B_{Deg}(0)$ over a finite sample volume. In contrast, adiabatic turn-off for collinear alignment of $B_{Deg}$ and $B_{Det}$ is correspondingly easy to accomplish if $B_{Deg} \parallel B_{Det}$ during the last cycle.

\section{Conclusion}
The type-I superconductor Pb shows significant flux rearrangement once it is driven into the intermediate state. However, excess low frequency noise after pulsed fields caused by rearrangement of flux within superconducting pick-up coils can be avoided by suitable a turn-off procedure. We used a linear decaying sinusoidal function and simulations showed that the spin-dynamics is strongly influenced during the degaussing process for $B_{Det}\perp B_{Deg}$. With this approach, we found that Nb can be degaussed within 50~ms and this is more easily achieved compared to Pb. For applications such as NCI or MEG-MRI, shorter degaussing times than 50~ms are desirable to minimize signal loss due to relaxation which can possibly be achieved in Nb wires.

\section*{Acknowledgment}

This work has received funding from the European Union's Horizon 2020 research and innovation programme under grant agreement No 686865 and by the DFG under grant No KO 5321/1-1.

\ifCLASSOPTIONcaptionsoff
  \newpage
\fi

\bibliographystyle{IEEEtran}
\bibliography{EUCAS_1EP3_08_FINAL_VERSION}

\end{document}